\documentclass[aps,prl,reprint]{revtex4-2}

\usepackage[T1]{fontenc}
\usepackage[american]{babel}
\usepackage{graphicx}
\usepackage{amsmath,amssymb,graphicx,bm,microtype} 
\usepackage[dvipsnames]{xcolor} 
\usepackage{booktabs} 
\usepackage[colorlinks,allcolors=blue!70!black]{hyperref} \usepackage[all]{hypcap} 
\usepackage[capitalise]{cleveref} 
\usepackage{orcidlink}
\newcommand{\orcid}[1]{\,\orcidlink{#1}}
\usepackage{braket}
\usepackage{siunitx,textcomp} 
\let\Re\relax \DeclareMathOperator{\Re}{Re}
\let\Im\relax \DeclareMathOperator{\Im}{Im}

\begin{document}

\title{Noise Reduction for Universal Hybrid Oscillator–Qubit Quantum Computation}

\author{Mohammad Nobakht\orcid{0009-0008-7016-1754}}
\email[Email: ]{nmnobakht@gmail.com}
\affiliation{School of Chemistry, University of Sydney, NSW 2006, Australia}
\author{Ivan Kassal\orcid{0000-0002-8376-0819}}
\email[Email: ]{ivan.kassal@sydney.edu.au}
\affiliation{School of Chemistry, University of Sydney, NSW 2006, Australia}

\begin{abstract}
Hybrid continuous-variable--discrete-variable (CV--DV) architectures process quantum information in bosonic modes and qubits, but noise limits their performance.
To reduce the noise, existing DV error correction must be complemented by CV noise reduction.
Existing CV noise-reduction schemes---such as GKP-stabilizer codes---can reduce CV noise, but only for Gaussian gates. Therefore, no current noise-reduction scheme can correct arbitrary CV--DV gates, including non-Gaussian ones.
Here, we develop noise reduction for a universal CV--DV gate set, making it applicable to arbitrary CV--DV gates.
We do so by introducing an ancilla qubit into a GKP-stabilizer code, allowing us to reduce the standard deviation of Gaussian displacement noise from \(\sigma\) to \(\tilde O(\sigma^2)\).
To demonstrate the scheme, we show that it significantly reduces noise and improves fidelity in the preparation of non-Gaussian cat and Fock states.
\end{abstract}

\maketitle
Hybrid continuous-variable--discrete-variable (CV--DV) architectures store quantum information in bosonic modes (CV) and qubits (DV)~\cite{andersen2015hybrid,liu2024hybrid,kemper2025hybrid,QSP_Steven}.
They exploit the complementary strengths of the two components to implement primitives that would be costly in purely CV or purely DV architectures.
For example, DV simulations of even elementary CV operations, such as displacements and beam splitters, require deep circuits with many entangling gates~\cite{crane2024hybrid,liu2024hybrid,nourse2025using}.
Conversely, ancilla DV degrees of freedom can enable non-Gaussian CV operations that would be costly when implemented using only CV resources~\cite{Lloyd1999,Braunstein2005,macdonell_analog_2021,Direct_estimation,navickas_experimental_2025,valahu_sensing_2025,chalermpusitarak2025programmable,Cubic2021Phase,Krastanov2015Universal,Conditional2024Not}.

Realizing this hybrid advantage requires reducing noise. While DV error correction is an advanced field~\cite{Shor,gottesman1997stabilizer,dennis2002topological,lidar2013quantum,Terhal,Girvin,ErrorCorrectionZoo}, CV noise reduction is less developed.
Many bosonic error-correcting codes, including the GKP, cat, binomial, and number-phase codes~\cite{Albert2025BosonicCoding,gottesman2001encoding,leghtas2013hardware,michael2016new,PhysRevA.97.032323,grimsmo2020quantum,PhysRevA.65.052316,wu2023optimal,PhysRevA.110.022402,terhal2020towards,Qubit_Oscillator_Concatenated}, protect information in finite-dimensional subspaces of a bosonic mode.
The encoded information therefore behaves as a qudit~\cite{brock_quantum_2025,joshi2021quantum,cai_bosonic_2021} and does not preserve the CV nature of the information.

Codes that preserve the CV nature of the information face two limitations.
First, codes that only use bosons cannot achieve a threshold~\cite{hanggli2022oscillator}, making noise reduction (as opposed to correction) the natural goal.
Second, existing codes are limited to noise reduction for Gaussian gates. 
They include GKP-stabilizer codes~\cite{noh2020encodingar,wu2023optimal,hanggli2022oscillator,brady2024safeguarding,Qubit_Oscillator_Concatenated}, which use non-Gaussian GKP ancillae to circumvent the Gaussian no-go theorem, which states that Gaussian resources alone cannot reduce Gaussian noise~\cite{PhysRevLett.102.120501,Distilling_Impossible}.
However, GKP-stabilizer codes can only reduce noise for Gaussian gates because they are the only gates that can be implemented logically as products of Gaussian operations acting separately on data and ancilla modes~\cite{noh2020encodingar}.
Therefore, no scheme can reduce noise in arbitrary CV or CV--DV gates. 

Here, we achieve noise reduction for a universal CV--DV gate set by introducing an ancilla qubit into a GKP-stabilizer code.
We exploit the fact that there are universal CV--DV gate sets that are linear in the CV quadratures, which enables them to be logically implemented within the GKP-stabilizer framework with operations that act separately on the data and ancilla modes.
To illustrate the utility of the logical gate set, we show that it increases the fidelity of cat-state and Fock-state preparation.

\begin{figure*}
    \centering
    \includegraphics[width=1\linewidth]{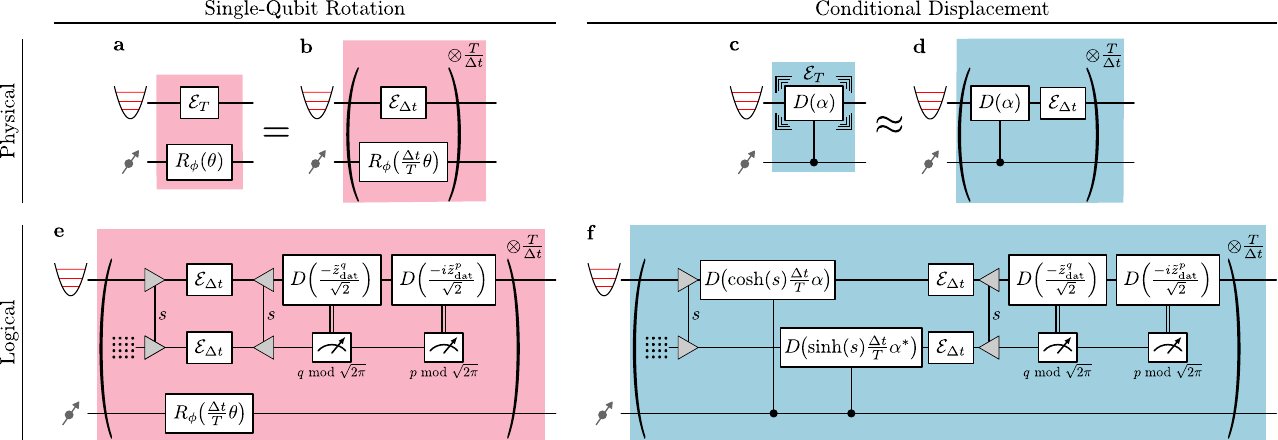}
\caption{\textbf{Physical and logical implementations of universal CV--DV gates under Gaussian displacement noise.}
\textbf{(a)}~Single-qubit rotation~(SQR) and noisy data-mode idling, modeled by Gaussian displacement channel \(\mathcal{E}_T\), where \(T\) is the gate duration. Parabola: data mode; arrow: qubit.
\textbf{(c)}~Noisy physical conditional displacement~(CD) gate.
\textbf{(b,d)}~Corresponding first-order Trotter discretizations into \(N=T/\Delta t\) time steps of duration \(\Delta t\), which decomposes \(\mathcal{E}_T\) into \(N\) noise channels \(\mathcal{E}_{\Delta t}\).
\textbf{(e,f)}~Corresponding logical implementations, using a GKP ancilla (shown as a lattice) and two-mode-squeezing encoder (two gray triangles) with squeezing parameter \(s\). The data-mode noise is corrected by counterdisplacement by the noise estimated from GKP syndrome measurements.
}
    \label{fig1}
\end{figure*}

\paragraph{Setup.}
We make three assumptions to isolate the intrinsic noise-reduction performance of our scheme. First, we use a code-capacity noise model with ideal encoding and decoding.
Second, we assume ideal GKP ancillae, neglecting imperfections in GKP-state preparation.
Finally, we assume access to an ideal ancilla qubit and ideal single-qubit gates, since DV error correction can be treated independently.

We consider a CV--DV system consisting of bosonic modes and qubits.
We assume access to single-qubit rotations~(SQR)
\begin{equation}
R_{\phi}(\theta)=\exp(-i\theta\sigma_{\phi}/2),
\end{equation}
where $\sigma_{\phi}=\sigma_{x}\cos\phi+\sigma_{y}\sin\phi$, and conditional displacements~(CD)
\begin{equation}
\label{eq:CD_def_alpha_single}
\mathrm{CD}(\alpha)
=
\exp[\sigma_z(\alpha a^\dagger-\alpha^\ast a)],
\end{equation}
where $a$ is the mode lowering operator and $\alpha\in\mathbb{C}$.
SQRs and CDs are universal for CV--DV systems because the dynamical Lie algebra generated by $\{\sigma_x,\sigma_y\}\cup\{\sigma_z\hat q,\sigma_z\hat p\}$ 
is sufficient for universal hybrid control~\cite{sutherland2021universal,eickbusch2022fast,liu2024hybrid,Conditional2024Not,kang_leveraging_2025}.

We model CV-mode noise as classical stochastic drives coupled linearly and independently to the position $\hat q$ and momentum $\hat p$, generating phase-space displacements. 
A single-mode displacement is \(D(\gamma)=\exp(\gamma a^\dagger-\gamma^\ast a)\).
For real \(\gamma\), \(D(\gamma)=\exp(-i\sqrt{2}\gamma\,\hat p)\) and \(D(i\gamma)=\exp(i\sqrt{2}\gamma\,\hat q)\), corresponding to \(\hat q\) and \(\hat p\) displacements, respectively.
Other physically relevant single-mode error channels, including loss and dephasing, admit displacement-operator representations~\cite{gottesman2001encoding,Knill_GKP,noh2018quantum,Albert2025BosonicCoding}. Therefore, a scheme that reduces displacement errors would suppress these channels as well, although the quantitative improvement would depend on how the channel weight is distributed over displacement amplitudes.

The displacement-noise Hamiltonian is
\begin{equation}
\label{eq_classical_stochastic_drive}
H_{\mathrm{noise}}(t)=f^q(t)\hat p - f^p(t)\hat q,
\end{equation}
where $f^\mu(t)$ for $\mu\in\{q,p\}$ are independent zero-mean Gaussian white-noise processes, 
$\mathbb{E}\left[f^\mu(t)f^\nu(t')\right]=\kappa\delta_{\mu\nu}\delta(t-t')$.
The noise accumulated over duration $T$ is
\begin{equation}
W^\mu(T)=\int_{0}^{T}dt\,f^\mu(t)\sim \mathcal{N}(0,\sigma_P^2),
\end{equation}
where $\sigma_P^2=\kappa T$ and $\mathcal{N}(\nu,\sigma^2)$ is a normal distribution with mean $\nu$ and variance $\sigma^2$.
During an SQR, $H_{\mathrm{noise}}(t)$ affects the idling CV mode only, giving the noise channel $\mathcal{E}_T$ in \cref{fig1}a. We model a noisy CD as evolution for time $T$ under
\begin{equation}
\label{Noisy_CD}
\tilde H_{\mathrm{CD}}(t)=H_{\mathrm{CD}}+H_{\mathrm{noise}}(t),
\end{equation}
where $H_{\mathrm{CD}}=i\sigma_z(\alpha a^\dagger-\alpha^\ast a)/T$ generates $\mathrm{CD}(\alpha)$.

\begin{figure*}
    \centering
    \includegraphics[width=1\linewidth]{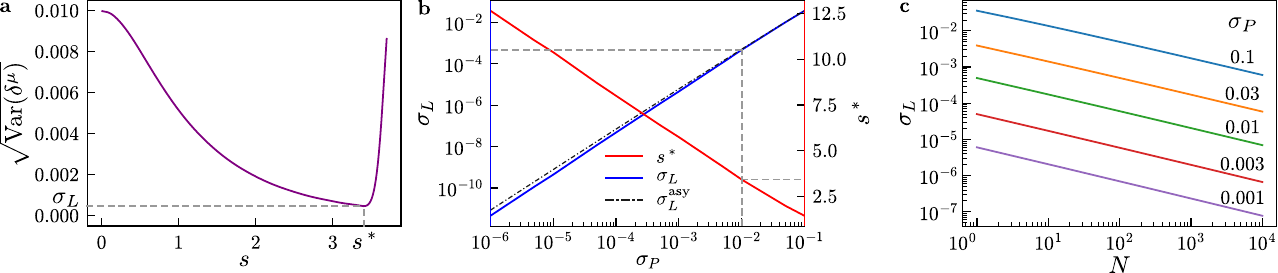}
\caption{\textbf{Logical noise for universal CV--DV gates.}
\textbf{(a)}~Standard deviation of each quadrature $\delta^\mu$ of the residual CV noise after decoding, on either an SQR or a CD gate, depends on the squeezing $s$. Example shown is for physical noise $\sigma_P=0.01$ and without Trotterization ($\sigma_P=\sigma_{P,N}$). The logical noise $\sigma_L$ is the minimum residual noise, obtained at optimal squeezing \(s^\ast\), calculated numerically.
\textbf{(b)}~The optimal squeezing \(s^\ast\) and the resulting logical noise \(\sigma_L\) as functions of \(\sigma_P\). The logical noise agrees with the asymptotic expression \(\sigma_L^\mathrm{asy}=\tilde O(\sigma_P^2)\), \cref{eq:asymptotic} (the seemingly larger discrepancy at small $\sigma_P$ is a log-scale artifact; \cref{eq:asymptotic} becomes more accurate as $\sigma_P\to 0$). No Trotterization is used; dotted lines indicate values from (a). 
\textbf{(c)}~Logical noise \(\sigma_{L}\), after Trotterization into $N$ steps, scales as \(\tilde O(\sigma_P^2/\sqrt{N})\).}
    \label{fig2}
\end{figure*}

We discretize the SQR and CD gates into $N$ short time steps of duration $\Delta t=T/N$ using a Trotter decomposition (\cref{fig1}b,d) for two reasons.
First, the noise-reduction scheme below yields stronger noise reduction for small noise, so discretizing the evolution reduces the total logical noise by keeping the noise in each time step small.
Second, up to Trotter error, it allows us to decompose the noisy CD gate into alternating ideal gate segments and pure noise channels.

During time step $j$ of a noisy SQR, the idle CV mode undergoes a random displacement by 
\(
\mathbf W_j(\Delta t)=(W^q_{j}(\Delta t), W^p_{j}(\Delta t))^{\top}
\), where the variance in each quadrature is \(\sigma_{P,N}^2=\kappa\Delta t\) in each time step.
The corresponding noise channel is \(\mathcal{E}_{\Delta t}\) in \cref{fig1}b.

For the noisy CD, the first-order Trotter splitting of the evolution under $\tilde H_{\mathrm{CD}}(t)$ over each time step gives an ideal CD followed by a random displacement induced by
 \(\mathcal{E}_{\Delta t}\) (\cref{fig1}d),
\begin{equation}
\mathcal{T}e^{
-i\int_{t_j}^{t_j+\Delta t}dt
\tilde H_{\mathrm{CD}}
}
\approx
e^{
-i W_j^q(\Delta t) \hat{p}+i W_j^p (\Delta t) \hat{q}}
e^{{-i H_{\mathrm{CD}}\Delta t}},
\end{equation}
The leading error term is $\tfrac12[A,B]$, where $A=-i W_j^q(\Delta t) \hat{p}+i W_j^p (\Delta t) \hat{q}$ and $B=-iH_{\mathrm{CD}}\Delta t$.
Since $\mathbf W_j(\Delta t)=O(\sqrt{\kappa\Delta t})$, the error is
$O(\sqrt{\kappa}\Delta t^{3/2}|\alpha|/T)$
per time step.
We therefore choose 
$N \gg (|\alpha| \sqrt{\kappa T})^{2/3}$ to keep the error small.

\paragraph{Logical CD gate.}
To reduce noise, we use the GKP--two-mode-squeezing (TMS) code~\cite{noh2020encodingar}, which encodes a data mode (``dat''), using a qunaught GKP ancilla~\cite{gottesman2001encoding} (``anc'').
The qunaught GKP ancilla is defined using the stabilizers
\(
\langle e^{i\sqrt{2\pi}\hat{q}},\, e^{-i\sqrt{2\pi}\hat{p}} \rangle
\),
chosen to maximize the range of noise that can be detected and corrected.
The Gaussian encoder of the TMS code is the TMS unitary
\begin{equation}
U_{\mathrm{enc}}=U_{\mathrm{TMS}}(s)=\exp\bigl[s(a_{\mathrm{dat}}^\dagger a_{\mathrm{anc}}^\dagger-a_{\mathrm{dat}} a_{\mathrm{anc}})\bigr],
\end{equation}
with squeezing $s>0$.
It acts on the quadrature vector $\hat{\mathbf r}=(\hat q_{\mathrm{dat}},\hat p_{\mathrm{dat}},\hat q_{\mathrm{anc}},\hat p_{\mathrm{anc}})^{\top}$ as the symplectic transformation
\begin{equation}
S_{\mathrm{TMS}}(s)
=
\begin{pmatrix}
\cosh (s) I & \sinh (s) Z\\
\sinh (s) Z & \cosh (s) I
\end{pmatrix},
\end{equation}
where $I=\mathrm{diag}(1,1)$, and $Z=\mathrm{diag}(1,-1)$.

To achieve universal noise reduction, we reduce noise in both SQRs and CDs. 
For SQRs, we use the existing TMS code to protect the data mode against idling noise (\cref{fig1}e).
Then, our main result is a protocol for reducing CD noise.  
Its five steps are illustrated in \cref{fig1}f: encoding, physical CD gates, noise, inverse encoding, and measurement and correction.
We denote the Heisenberg-picture quadrature vector after step $i$ by $\hat{\mathbf r}^{(i)}$ and, from now on, omit the time-step subscript $j$ on noise random variables.

\paragraph*{Step~1: Encoding.}
The encoded logical state is
\begin{equation}
\label{eq:tms_encoding}
\ket{\psi_L}
=
U_{\mathrm{TMS}}(s)\bigl(\ket{\psi}_{\mathrm{dat}}\otimes\ket{\mathrm{GKP}}_{\mathrm{anc}}\bigr),
\end{equation}
with quadrature vector
\begin{equation}
\label{eq:tms_heisenberg}
\hat{\mathbf r}^{(1)}=U_{\mathrm{TMS}}^{\dagger}(s)\hat{\mathbf r}U_{\mathrm{TMS}}(s)=S_{\mathrm{TMS}}(s)\hat{\mathbf r}.
\end{equation}

\paragraph*{Step~2: Physical CD gates.}
During each time step, we apply CD gates on both modes,
\begin{equation}
\mathrm{CD}_{\mathrm{anc}}(\sinh (s)\alpha^\ast /N)\,
\mathrm{CD}_{\mathrm{dat}}(\cosh (s)\alpha /N),
\end{equation}
inducing the qubit-conditional displacement to
\begin{equation}
\label{eq:Step2_Heis_compact}
\hat{\mathbf r}^{(2)}
=
\hat{\mathbf r}^{(1)}
+\sigma_z\sqrt{2}\,\boldsymbol{\xi}, \quad \boldsymbol{\xi}
=
\frac{1}{N}
\begin{pmatrix}
\cosh (s)I\bm\alpha_{\mathbb R}\\
\sinh (s)Z\bm\alpha_{\mathbb R}
\end{pmatrix},
\end{equation}
where $\bm\alpha_{\mathbb R}=(\Re\alpha,\Im\alpha)^{\top}$.

\paragraph*{Step~3: Noise.}
$H_{\mathrm{noise}}(t)$ acts independently on both modes, displacing them by
\begin{equation}
\mathbf \Omega=(\mathbf W_{{\mathrm{dat}}}(\Delta t), \mathbf W_{\mathrm{anc}}(\Delta t))^{\top},
\end{equation}
inducing the update
\begin{align}
\hat{\mathbf r}^{(3)}
& =
\hat{\mathbf r}^{(2)}+\mathbf \Omega
=S_{\mathrm{TMS}}(s)\hat{\mathbf r}
+
\sigma_z\sqrt{2}\,\boldsymbol{\xi}
+
\mathbf \Omega.
\end{align}

\paragraph*{Step~4: Inverse encoding.}
Decoding starts with the inverse encoding unitary $U_{\mathrm{enc}}^{\dagger}$, giving
\begin{equation}
\label{eq:heis_after_decode}
\hat{\mathbf r}^{(4)}
=
S_{\mathrm{TMS}}^{-1}(s)\hat{\mathbf r}^{(3)}
=
\hat{\mathbf r}
+
\sigma_z\sqrt{2}\,\boldsymbol{\xi}_L
+
\bm z,
\end{equation}
where the logical displacement is $\boldsymbol{\xi}_L = (\bm\alpha_{\mathbb R},\mathbf{0})^{\top}/N$
and the transformed noise is
\begin{equation}
\label{eq:zj_def}
\bm z=S_{\mathrm{TMS}}^{-1}(s)\mathbf \Omega.
\end{equation}
Thus, we recover the target gate $\mathrm{CD}_\mathrm{dat}(\alpha/N)$, while transforming the independent noise $\mathbf \Omega$ into the correlated noise $\bm z$.
Because $\mathbb{E}[\mathbf \Omega]=\bm 0$ and $\mathrm{Cov}(\mathbf \Omega)=\kappa\Delta tI_4$, we have $\mathbb{E}[\bm z]=\bm 0$ and
\begin{align}
\mathrm{Cov}(\bm z)
&=\kappa\Delta t \, S_{\mathrm{TMS}}^{-1}(s)S_{\mathrm{TMS}}^{-\top}(s) \\ 
&=\kappa\Delta t
\begin{pmatrix}
\cosh(2s)I & -\sinh(2s)Z \\
-\sinh(2s)Z & \cosh(2s)I
\end{pmatrix}. \label{eq:cov-matrix}
\end{align}
The off-diagonal blocks show that the TMS code induces correlations between the noise on the two modes, growing as \(\sinh(2s)\) with the squeezing \(s\).

\begin{figure*}
    \centering
    \includegraphics[width=1\linewidth]{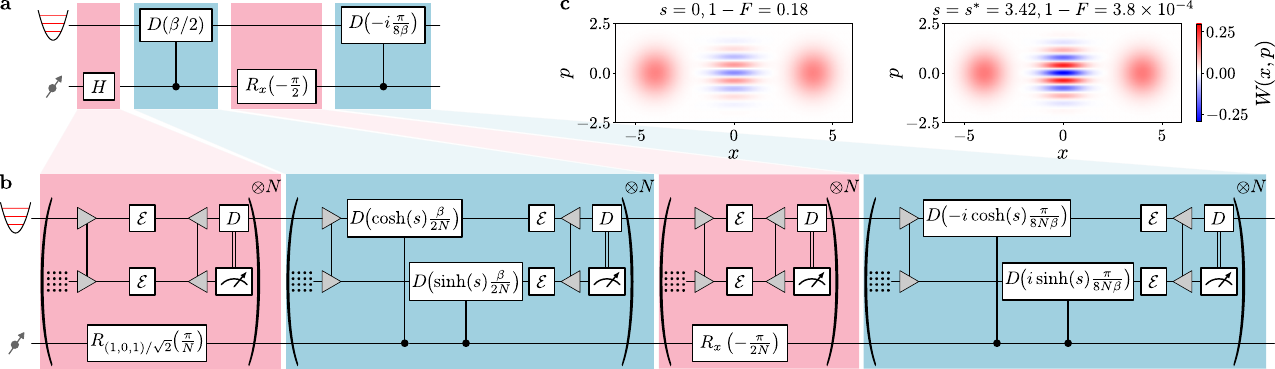}
\caption{\textbf{Noise reduction for cat-state preparation.}
\textbf{(a)}~Circuit for preparing an approximation to the cat state
$\ket{C_-(\beta)} \propto \ket{\beta/\sqrt{2}} - \ket{-\beta/\sqrt{2}}$~\cite{girvin2019schrodinger}.
\textbf{(b)}~
Logical circuit obtained by splitting each physical gate in (a) into \(N\) time steps, with each time step replaced by its logical implementation from \cref{fig1}. \(\mathcal{E}\) denotes \(\mathcal{E}_{\Delta t}\) and each correction is depicted as a single measurement and displacement.
\textbf{(c)}~Output Wigner functions for the physical (\(s=0\)) and logical (\(s=s^\ast\)) implementations.
The logical implementation reduces phase-space blurring, preserves Wigner negativity, and decreases the average infidelity \(1-F\) with respect to the noiseless circuit from 0.18 to $3.8\times10^{-4}$.
Parameters: \(\beta=4\), \(\sigma_P=0.1\), \(N=100\), and \(s^\ast(\sigma_{P,N})=3.42\).
}
    \label{fig3}
\end{figure*}

\paragraph*{Step~5: Measurement and correction.} 
We estimate the data-mode noise from GKP syndrome measurements on the ancilla mode, which give 
\begin{equation}
    \label{eq:ancilla_reduced}
    \bar {\bm z}_{\mathrm{anc}}=\bm z_{\mathrm{anc}}\!\!\! \mod \sqrt{2\pi},
\end{equation}
where the outcome for each component is in \([-\sqrt{2\pi}/2,\sqrt{2\pi}/2)\), the GKP correctable window~\cite{gottesman2001encoding}.

Using the correlations in \cref{eq:cov-matrix}, we estimate the data-mode noise in each quadrature from the ancilla syndrome measurements. Our estimate assumes that the ancilla noise lies within the GKP correctable window, i.e., that the modular outcome \(\bar {\bm z}_{\mathrm{anc}}\) equals the true ancilla noise \(\bm z_{\mathrm{anc}}\). Then, the linear minimum mean-square error (LMMSE) estimator~\cite[Sec.~12.3]{Kay1993} is
\begin{align}
\tilde {\bm z}_{\mathrm{dat}}
&=
\mathrm{Cov}(\bm z_{\mathrm{dat}}, \bm z_{\mathrm{anc}})\, \mathrm{Cov}(\bm z_{\mathrm{anc}})^{-1} \bar {\bm z}_{\mathrm{anc}} \\
&=
-Z\tanh(2s)\bar {\bm z}_{\mathrm{anc}}.
\end{align}
We then apply the counter-displacement correction $-\tilde {\bm z}_{\mathrm{dat}}$, i.e., 
\(D(-\tilde z_{\mathrm{dat}}^{q}/\sqrt{2})
D(-i\tilde z_{\mathrm{dat}}^{p}/\sqrt{2})\),
to the data mode to remove the estimated noise.
The residual noise is then
\begin{equation}
\label{eq:logical_output_noise}
\bm\delta  = \bm z_{\mathrm{dat}} - \tilde {\bm z}_{\mathrm{dat}}.\end{equation}

We set $s$ to minimize \(\mathrm{Var}(\delta^\mu)\), which is equal for both quadratures.
If the true (non-modular) ancilla noise \(\bm z_{\mathrm{anc}}\) were accessible, \(\mathrm{Var}(\delta^\mu)\) would be
\(
\kappa \Delta t\operatorname{sech}(2s),
\)
which would improve noise reduction exponentially with \(s\).
However, increasing \(s\) also increases each \(\mathrm{Var}(z^\mu_{\mathrm{anc}})\) to
\(
\kappa \Delta t \cosh(2s)
\).
Therefore, as $s$ is increased, there is greater probability that the ancilla noise will not remain within the GKP correctable window, leading to additional error because the modular measurement outcomes differ from the true ancilla noise.
There is therefore a finite optimal squeezing \(s^*\) that minimizes \(\mathrm{Var}(\delta^\mu)\), which we determine numerically in \cref{fig2}a.

Finally, we define the logical noise per time step to be \(\bm \delta(s^*)\), whose quadratures have equal standard deviation \(\sigma_{L,N} = \min_s \mathrm{Var}(\delta^\mu)^{1/2}\).

\paragraph{Performance.}
When the Trotter number $N$ is large enough that the Trotter error is negligible and the physical noise per time step is small, $\sigma_{P,N}=\sigma_P/\sqrt{N}\ll 1$, our scheme reduces the data-mode noise over full SQRs and CDs from $\sigma_P$ to $\sigma_L=\tilde O(\sigma_P^2/\sqrt{N})$, where $\tilde O$ neglects logarithmic factors. In this regime, the TMS code for idling CV information, and hence for our SQR, gives the asymptotic per-step logical noise~\cite{noh2020encodingar}
\begin{equation}
    \sigma^{\mathrm{asy}}_{L,N}=\frac{2\sigma_{P,N}^2}{\sqrt{\pi}}\ln^{1/2}\!\bigg(\frac{\pi^{3/2}}{2\sigma_{P,N}^4}\bigg).
    \label{eq:asymptotic}
\end{equation}
The same per-step logical noise also applies to our encoded CD gate, because the transformed noise $\bm z$ in \crefrange{eq:zj_def}{eq:cov-matrix} is independent of $\alpha$; equivalently, setting $\alpha=0$ reduces the encoded CD to idling, as shown in \cref{fig2}b. Because the logical-noise contributions from different time steps are independent, the full-gate logical noise is 
$\sigma_L=\sqrt{N}\sigma_{L,N}=\widetilde O(\sigma_P^2/\sqrt{N})$,
as shown in \cref{fig2}c.

\begin{figure*}
    \centering
    \includegraphics[width=1\linewidth]{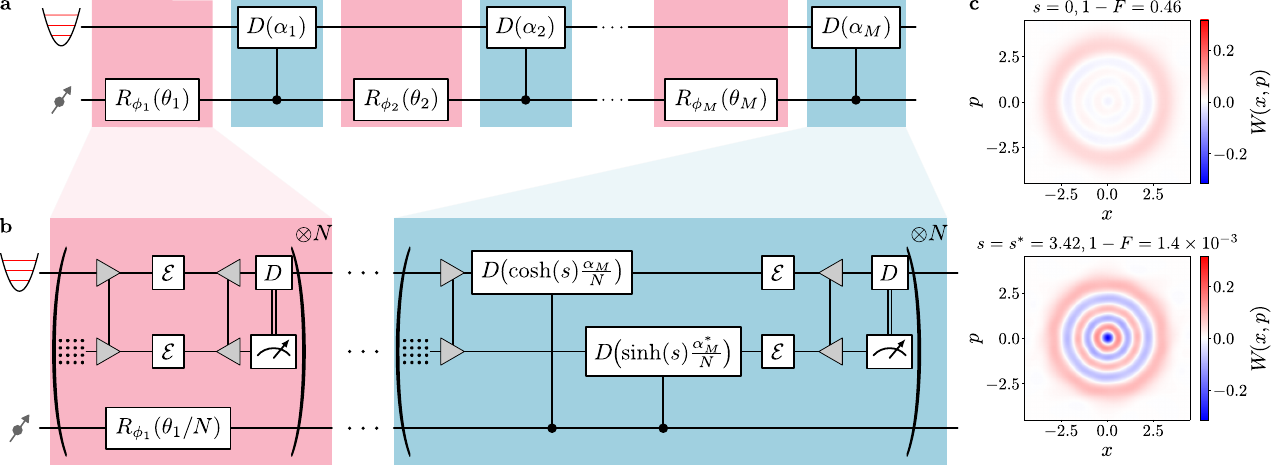}
\caption{\textbf{Noise reduction for Fock-state preparation.}
\textbf{(a)}~Circuit with output approximating \(\ket{n=5}\), using optimized parameters from~\cite{eickbusch2022fast}.
\textbf{(b)}~Logical circuit obtained by splitting each physical gate in (a) into \(N\) time steps, with each time step replaced by its logical implementation from \cref{fig1}.
\textbf{(c)}~Output Wigner functions for the physical (\(s=0\)) and logical (\(s=s^\ast\)) implementations.
The logical implementation reduces phase-space blurring, preserves Wigner negativity, and decreases the average infidelity \(1-F\) with respect to the noiseless circuit from 0.46 to \num{1.4e-3}.
Parameters: \(M=9\), \(\sigma_P=0.1\), \(N=100\), and \(s^\ast(\sigma_{P,N})=3.42\).}
\label{fig4}
\end{figure*}

\textit{Examples.}
We give two examples of non-Gaussian state preparation to illustrate the significant fidelity improvement enabled by our approach: cat-state preparation in \cref{fig3} and Fock-state preparation in \cref{fig4}.
In both examples, we decompose the target circuit into gates from the universal gate set and then replace each gate by its logical implementation.
We quantify the performance by the state infidelity $1-F$, averaged over the noise distribution.
In both cases, the logical implementation yields higher average fidelity and better preserves Wigner negativity than the physical implementation.

\paragraph{Discussion.}
We showed that, when \(N\) is chosen large enough that \(\sigma_{P,N}\ll 1\), the logical noise of both SQRs and CDs scales as \(\tilde O(\sigma_P^2/\sqrt{N})\). Therefore, our results establish noise reduction for a universal CV--DV gate set, with the same noise-reduction performance as the TMS code achieves when protecting idling CV information~\cite{noh2020encodingar}.
As examples, we used our approach to prepare cat and Fock states with higher fidelity.

Two directions could improve and extend our approach.
First, replacing the qunaught GKP ancilla with an oscillator--qudit locally compact abelian (LCA) state of dimension \(c\) would increase the correctable window from \(\sqrt{2\pi}\) to \(\sqrt{2\pi c}\)~\cite{Sayan}, thus improving the achievable noise reduction.
Second, our noise-reduction scheme extends to any GKP-stabilizer code. Every GKP-stabilizer code is defined by a Gaussian encoding unitary. Conjugation by this unitary maps any logical single-mode CD gate to a product of physical CD gates, where the shift amplitudes are determined by the encoding. Therefore, better noise reduction could be achieved by using, for example, GKP-stabilizer codes that use additional GKP ancillas~\cite{brady2024safeguarding}. However, in both cases, the performance improvements could involve tradeoffs with the cost of additional resources or more complicated encodings. 

Our analysis treated as negligible three sources of error that could play a role in practical implementations: imperfect GKP ancillae, noise in the encoding and decoding, and Trotter error. Under these assumptions, applying noise reduction at each of \(N\) Trotter steps yields less accumulated logical noise than applying it only once after the full gate, and this improvement increases with \(N\). However, increasing \(N\) also requires more GKP ancillae and more encoding and decoding steps. With imperfect GKP ancillae and noisy encoding and decoding, this tradeoff is expected to yield an optimal finite \(N\), which could be evaluated on a case-by-case basis.

Overall, our work establishes noise reduction for a universal CV--DV gate set and opens a path to more reliable CV--DV quantum information processing.

\begin{acknowledgments}
We thank Andrew Doherty, Timo Hillmann, Sahand Mahmoodian, Kai Schwennicke, and Ting Rei Tan for valuable discussions. We were supported by the Sydney Quantum Academy and by the Australian Research Council (FT230100653).
\end{acknowledgments}

\bibliography{bib}

\begin{thebibliography}{50}%
\makeatletter
\providecommand \@ifxundefined [1]{%
 \@ifx{#1\undefined}
}%
\providecommand \@ifnum [1]{%
 \ifnum #1\expandafter \@firstoftwo
 \else \expandafter \@secondoftwo
 \fi
}%
\providecommand \@ifx [1]{%
 \ifx #1\expandafter \@firstoftwo
 \else \expandafter \@secondoftwo
 \fi
}%
\providecommand \natexlab [1]{#1}%
\providecommand \enquote  [1]{``#1''}%
\providecommand \bibnamefont  [1]{#1}%
\providecommand \bibfnamefont [1]{#1}%
\providecommand \citenamefont [1]{#1}%
\providecommand \href@noop [0]{\@secondoftwo}%
\providecommand \href [0]{\begingroup \@sanitize@url \@href}%
\providecommand \@href[1]{\@@startlink{#1}\@@href}%
\providecommand \@@href[1]{\endgroup#1\@@endlink}%
\providecommand \@sanitize@url [0]{\catcode `\\12\catcode `\$12\catcode `\&12\catcode `\#12\catcode `\^12\catcode `\_12\catcode `\%12\relax}%
\providecommand \@@startlink[1]{}%
\providecommand \@@endlink[0]{}%
\providecommand \url  [0]{\begingroup\@sanitize@url \@url }%
\providecommand \@url [1]{\endgroup\@href {#1}{\urlprefix }}%
\providecommand \urlprefix  [0]{URL }%
\providecommand \Eprint [0]{\href }%
\providecommand \doibase [0]{https://doi.org/}%
\providecommand \selectlanguage [0]{\@gobble}%
\providecommand \bibinfo  [0]{\@secondoftwo}%
\providecommand \bibfield  [0]{\@secondoftwo}%
\providecommand \translation [1]{[#1]}%
\providecommand \BibitemOpen [0]{}%
\providecommand \bibitemStop [0]{}%
\providecommand \bibitemNoStop [0]{.\EOS\space}%
\providecommand \EOS [0]{\spacefactor3000\relax}%
\providecommand \BibitemShut  [1]{\csname bibitem#1\endcsname}%
\let\auto@bib@innerbib\@empty
\bibitem [{\citenamefont {Andersen}\ \emph {et~al.}(2015)\citenamefont {Andersen}, \citenamefont {Neergaard-Nielsen}, \citenamefont {van Loock},\ and\ \citenamefont {Furusawa}}]{andersen2015hybrid}%
  \BibitemOpen
  \bibfield  {author} {\bibinfo {author} {\bibfnamefont {U.~L.}\ \bibnamefont {Andersen}}, \bibinfo {author} {\bibfnamefont {J.~S.}\ \bibnamefont {Neergaard-Nielsen}}, \bibinfo {author} {\bibfnamefont {P.}~\bibnamefont {van Loock}},\ and\ \bibinfo {author} {\bibfnamefont {A.}~\bibnamefont {Furusawa}},\ }\bibfield  {title} {\bibinfo {title} {Hybrid discrete- and continuous-variable quantum information},\ }\href {https://doi.org/10.1038/nphys3410} {\bibfield  {journal} {\bibinfo  {journal} {Nat. Phys.}\ }\textbf {\bibinfo {volume} {11}},\ \bibinfo {pages} {713} (\bibinfo {year} {2015})}\BibitemShut {NoStop}%
\bibitem [{\citenamefont {Liu}\ \emph {et~al.}(2026)\citenamefont {Liu}, \citenamefont {Singh}, \citenamefont {Smith}, \citenamefont {Crane}, \citenamefont {Martyn}, \citenamefont {Eickbusch}, \citenamefont {Schuckert}, \citenamefont {Li}, \citenamefont {Sinanan-Singh}, \citenamefont {Soley}, \citenamefont {Tsunoda}, \citenamefont {Chuang}, \citenamefont {Wiebe},\ and\ \citenamefont {Girvin}}]{liu2024hybrid}%
  \BibitemOpen
  \bibfield  {author} {\bibinfo {author} {\bibfnamefont {Y.}~\bibnamefont {Liu}}, \bibinfo {author} {\bibfnamefont {S.}~\bibnamefont {Singh}}, \bibinfo {author} {\bibfnamefont {K.~C.}\ \bibnamefont {Smith}}, \bibinfo {author} {\bibfnamefont {E.}~\bibnamefont {Crane}}, \bibinfo {author} {\bibfnamefont {J.~M.}\ \bibnamefont {Martyn}}, \bibinfo {author} {\bibfnamefont {A.}~\bibnamefont {Eickbusch}}, \bibinfo {author} {\bibfnamefont {A.}~\bibnamefont {Schuckert}}, \bibinfo {author} {\bibfnamefont {R.~D.}\ \bibnamefont {Li}}, \bibinfo {author} {\bibfnamefont {J.}~\bibnamefont {Sinanan-Singh}}, \bibinfo {author} {\bibfnamefont {M.~B.}\ \bibnamefont {Soley}}, \bibinfo {author} {\bibfnamefont {T.}~\bibnamefont {Tsunoda}}, \bibinfo {author} {\bibfnamefont {I.~L.}\ \bibnamefont {Chuang}}, \bibinfo {author} {\bibfnamefont {N.}~\bibnamefont {Wiebe}},\ and\ \bibinfo {author} {\bibfnamefont {S.~M.}\ \bibnamefont {Girvin}},\ }\bibfield  {title} {\bibinfo {title} {Hybrid oscillator-qubit quantum processors: Instruction set
  architectures, abstract machine models, and applications},\ }\href {https://doi.org/10.1103/4rf7-9tfx} {\bibfield  {journal} {\bibinfo  {journal} {PRX Quantum}\ }\textbf {\bibinfo {volume} {7}},\ \bibinfo {pages} {010201} (\bibinfo {year} {2026})}\BibitemShut {NoStop}%
\bibitem [{\citenamefont {Kemper}\ \emph {et~al.}(2025)\citenamefont {Kemper}, \citenamefont {Alvertis}, \citenamefont {Asaduzzaman}, \citenamefont {Bakalov}, \citenamefont {Baron}, \citenamefont {Bierman}, \citenamefont {Burgstahler}, \citenamefont {Chundury}, \citenamefont {Das}, \citenamefont {Furches} \emph {et~al.}}]{kemper2025hybrid}%
  \BibitemOpen
  \bibfield  {author} {\bibinfo {author} {\bibfnamefont {A.~F.}\ \bibnamefont {Kemper}}, \bibinfo {author} {\bibfnamefont {A.}~\bibnamefont {Alvertis}}, \bibinfo {author} {\bibfnamefont {M.}~\bibnamefont {Asaduzzaman}}, \bibinfo {author} {\bibfnamefont {B.~N.}\ \bibnamefont {Bakalov}}, \bibinfo {author} {\bibfnamefont {D.}~\bibnamefont {Baron}}, \bibinfo {author} {\bibfnamefont {J.}~\bibnamefont {Bierman}}, \bibinfo {author} {\bibfnamefont {B.}~\bibnamefont {Burgstahler}}, \bibinfo {author} {\bibfnamefont {S.}~\bibnamefont {Chundury}}, \bibinfo {author} {\bibfnamefont {E.~R.}\ \bibnamefont {Das}}, \bibinfo {author} {\bibfnamefont {J.}~\bibnamefont {Furches}}, \emph {et~al.},\ }\href@noop {} {\bibinfo {title} {Hybrid continuous-discrete-variable quantum computing: A guide to utility}} (\bibinfo {year} {2025}),\ \Eprint {https://arxiv.org/abs/2511.13882} {arXiv:2511.13882} \BibitemShut {NoStop}%
\bibitem [{\citenamefont {Liu}\ \emph {et~al.}(2025)\citenamefont {Liu}, \citenamefont {Martyn}, \citenamefont {Sinanan-Singh}, \citenamefont {Smith}, \citenamefont {Girvin},\ and\ \citenamefont {Chuang}}]{QSP_Steven}%
  \BibitemOpen
  \bibfield  {author} {\bibinfo {author} {\bibfnamefont {Y.}~\bibnamefont {Liu}}, \bibinfo {author} {\bibfnamefont {J.~M.}\ \bibnamefont {Martyn}}, \bibinfo {author} {\bibfnamefont {J.}~\bibnamefont {Sinanan-Singh}}, \bibinfo {author} {\bibfnamefont {K.~C.}\ \bibnamefont {Smith}}, \bibinfo {author} {\bibfnamefont {S.~M.}\ \bibnamefont {Girvin}},\ and\ \bibinfo {author} {\bibfnamefont {I.~L.}\ \bibnamefont {Chuang}},\ }\bibfield  {title} {\bibinfo {title} {Toward mixed analog-digital quantum signal processing: Quantum {AD/DA} conversion and the {F}ourier transform},\ }\href {https://doi.org/10.1109/TSP.2025.3599462} {\bibfield  {journal} {\bibinfo  {journal} {IEEE Trans. Signal Process.}\ }\textbf {\bibinfo {volume} {73}},\ \bibinfo {pages} {3641} (\bibinfo {year} {2025})}\BibitemShut {NoStop}%
\bibitem [{\citenamefont {Crane}\ \emph {et~al.}(2024)\citenamefont {Crane}, \citenamefont {Smith}, \citenamefont {Tomesh}, \citenamefont {Eickbusch}, \citenamefont {Martyn}, \citenamefont {K{\"u}hn}, \citenamefont {Funcke}, \citenamefont {DeMarco}, \citenamefont {Chuang},\ and\ \citenamefont {Wiebe}}]{crane2024hybrid}%
  \BibitemOpen
  \bibfield  {author} {\bibinfo {author} {\bibfnamefont {E.}~\bibnamefont {Crane}}, \bibinfo {author} {\bibfnamefont {K.~C.}\ \bibnamefont {Smith}}, \bibinfo {author} {\bibfnamefont {T.}~\bibnamefont {Tomesh}}, \bibinfo {author} {\bibfnamefont {A.}~\bibnamefont {Eickbusch}}, \bibinfo {author} {\bibfnamefont {J.~M.}\ \bibnamefont {Martyn}}, \bibinfo {author} {\bibfnamefont {S.}~\bibnamefont {K{\"u}hn}}, \bibinfo {author} {\bibfnamefont {L.}~\bibnamefont {Funcke}}, \bibinfo {author} {\bibfnamefont {M.~A.}\ \bibnamefont {DeMarco}}, \bibinfo {author} {\bibfnamefont {I.~L.}\ \bibnamefont {Chuang}},\ and\ \bibinfo {author} {\bibfnamefont {N.}~\bibnamefont {Wiebe}},\ }\href@noop {} {\bibinfo {title} {Hybrid oscillator-qubit quantum processors: Simulating fermions, bosons, and gauge fields}} (\bibinfo {year} {2024}),\ \Eprint {https://arxiv.org/abs/2409.03747} {arXiv:2409.03747} \BibitemShut {NoStop}%
\bibitem [{\citenamefont {Nourse}\ \emph {et~al.}(2025)\citenamefont {Nourse}, \citenamefont {Olaya-Agudelo},\ and\ \citenamefont {Kassal}}]{nourse2025using}%
  \BibitemOpen
  \bibfield  {author} {\bibinfo {author} {\bibfnamefont {H.~L.}\ \bibnamefont {Nourse}}, \bibinfo {author} {\bibfnamefont {V.~C.}\ \bibnamefont {Olaya-Agudelo}},\ and\ \bibinfo {author} {\bibfnamefont {I.}~\bibnamefont {Kassal}},\ }\href@noop {} {\bibinfo {title} {Using bosons to improve resource efficiency of quantum simulation of vibronic molecular dynamics}} (\bibinfo {year} {2025}),\ \Eprint {https://arxiv.org/abs/2512.20828} {arXiv:2512.20828} \BibitemShut {NoStop}%
\bibitem [{\citenamefont {Lloyd}\ and\ \citenamefont {Braunstein}(1999)}]{Lloyd1999}%
  \BibitemOpen
  \bibfield  {author} {\bibinfo {author} {\bibfnamefont {S.}~\bibnamefont {Lloyd}}\ and\ \bibinfo {author} {\bibfnamefont {S.~L.}\ \bibnamefont {Braunstein}},\ }\bibfield  {title} {\bibinfo {title} {Quantum computation with continuous variables},\ }\href {https://doi.org/10.1103/PhysRevLett.82.1784} {\bibfield  {journal} {\bibinfo  {journal} {Phys. Rev. Lett.}\ }\textbf {\bibinfo {volume} {82}},\ \bibinfo {pages} {1784} (\bibinfo {year} {1999})}\BibitemShut {NoStop}%
\bibitem [{\citenamefont {Braunstein}\ and\ \citenamefont {van Loock}(2005)}]{Braunstein2005}%
  \BibitemOpen
  \bibfield  {author} {\bibinfo {author} {\bibfnamefont {S.~L.}\ \bibnamefont {Braunstein}}\ and\ \bibinfo {author} {\bibfnamefont {P.}~\bibnamefont {van Loock}},\ }\bibfield  {title} {\bibinfo {title} {Quantum information with continuous variables},\ }\href {https://doi.org/10.1103/RevModPhys.77.513} {\bibfield  {journal} {\bibinfo  {journal} {Rev. Mod. Phys.}\ }\textbf {\bibinfo {volume} {77}},\ \bibinfo {pages} {513} (\bibinfo {year} {2005})}\BibitemShut {NoStop}%
\bibitem [{\citenamefont {MacDonell}\ \emph {et~al.}(2021)\citenamefont {MacDonell}, \citenamefont {Dickerson}, \citenamefont {Birch}, \citenamefont {Kumar}, \citenamefont {Edmunds}, \citenamefont {Biercuk}, \citenamefont {Hempel},\ and\ \citenamefont {Kassal}}]{macdonell_analog_2021}%
  \BibitemOpen
  \bibfield  {author} {\bibinfo {author} {\bibfnamefont {R.~J.}\ \bibnamefont {MacDonell}}, \bibinfo {author} {\bibfnamefont {C.~E.}\ \bibnamefont {Dickerson}}, \bibinfo {author} {\bibfnamefont {C.~J.~T.}\ \bibnamefont {Birch}}, \bibinfo {author} {\bibfnamefont {A.}~\bibnamefont {Kumar}}, \bibinfo {author} {\bibfnamefont {C.~L.}\ \bibnamefont {Edmunds}}, \bibinfo {author} {\bibfnamefont {M.~J.}\ \bibnamefont {Biercuk}}, \bibinfo {author} {\bibfnamefont {C.}~\bibnamefont {Hempel}},\ and\ \bibinfo {author} {\bibfnamefont {I.}~\bibnamefont {Kassal}},\ }\bibfield  {title} {\bibinfo {title} {Analog quantum simulation of chemical dynamics},\ }\href {https://doi.org/10.1039/D1SC02142G} {\bibfield  {journal} {\bibinfo  {journal} {Chem. Sci.}\ }\textbf {\bibinfo {volume} {12}},\ \bibinfo {pages} {9794} (\bibinfo {year} {2021})}\BibitemShut {NoStop}%
\bibitem [{\citenamefont {Krisnanda}\ \emph {et~al.}(2026)\citenamefont {Krisnanda}, \citenamefont {Valadares}, \citenamefont {Chu}, \citenamefont {Song}, \citenamefont {Copetudo}, \citenamefont {Fontaine}, \citenamefont {Lachman}, \citenamefont {Filip},\ and\ \citenamefont {Gao}}]{Direct_estimation}%
  \BibitemOpen
  \bibfield  {author} {\bibinfo {author} {\bibfnamefont {T.}~\bibnamefont {Krisnanda}}, \bibinfo {author} {\bibfnamefont {F.}~\bibnamefont {Valadares}}, \bibinfo {author} {\bibfnamefont {K.~T.~N.}\ \bibnamefont {Chu}}, \bibinfo {author} {\bibfnamefont {P.}~\bibnamefont {Song}}, \bibinfo {author} {\bibfnamefont {A.}~\bibnamefont {Copetudo}}, \bibinfo {author} {\bibfnamefont {C.~Y.}\ \bibnamefont {Fontaine}}, \bibinfo {author} {\bibfnamefont {L.}~\bibnamefont {Lachman}}, \bibinfo {author} {\bibfnamefont {R.}~\bibnamefont {Filip}},\ and\ \bibinfo {author} {\bibfnamefont {Y.~Y.}\ \bibnamefont {Gao}},\ }\bibfield  {title} {\bibinfo {title} {Direct estimation of arbitrary observables of an oscillator},\ }\href {https://doi.org/10.1103/72f2-tgwp} {\bibfield  {journal} {\bibinfo  {journal} {Phys. Rev. Res.}\ }\textbf {\bibinfo {volume} {8}},\ \bibinfo {pages} {023026} (\bibinfo {year} {2026})}\BibitemShut {NoStop}%
\bibitem [{\citenamefont {Navickas}\ \emph {et~al.}(2025)\citenamefont {Navickas}, \citenamefont {MacDonell}, \citenamefont {Valahu}, \citenamefont {Olaya-Agudelo}, \citenamefont {Scuccimarra}, \citenamefont {Millican}, \citenamefont {Matsos}, \citenamefont {Nourse}, \citenamefont {Rao}, \citenamefont {Biercuk}, \citenamefont {Hempel}, \citenamefont {Kassal},\ and\ \citenamefont {Tan}}]{navickas_experimental_2025}%
  \BibitemOpen
  \bibfield  {author} {\bibinfo {author} {\bibfnamefont {T.}~\bibnamefont {Navickas}}, \bibinfo {author} {\bibfnamefont {R.~J.}\ \bibnamefont {MacDonell}}, \bibinfo {author} {\bibfnamefont {C.~H.}\ \bibnamefont {Valahu}}, \bibinfo {author} {\bibfnamefont {V.~C.}\ \bibnamefont {Olaya-Agudelo}}, \bibinfo {author} {\bibfnamefont {F.}~\bibnamefont {Scuccimarra}}, \bibinfo {author} {\bibfnamefont {M.~J.}\ \bibnamefont {Millican}}, \bibinfo {author} {\bibfnamefont {V.~G.}\ \bibnamefont {Matsos}}, \bibinfo {author} {\bibfnamefont {H.~L.}\ \bibnamefont {Nourse}}, \bibinfo {author} {\bibfnamefont {A.~D.}\ \bibnamefont {Rao}}, \bibinfo {author} {\bibfnamefont {M.~J.}\ \bibnamefont {Biercuk}}, \bibinfo {author} {\bibfnamefont {C.}~\bibnamefont {Hempel}}, \bibinfo {author} {\bibfnamefont {I.}~\bibnamefont {Kassal}},\ and\ \bibinfo {author} {\bibfnamefont {T.~R.}\ \bibnamefont {Tan}},\ }\bibfield  {title} {\bibinfo {title} {Experimental quantum simulation of chemical dynamics},\ }\href
  {https://doi.org/10.1021/jacs.5c03336} {\bibfield  {journal} {\bibinfo  {journal} {J. Am. Chem. Soc.}\ }\textbf {\bibinfo {volume} {147}},\ \bibinfo {pages} {23566} (\bibinfo {year} {2025})}\BibitemShut {NoStop}%
\bibitem [{\citenamefont {Valahu}\ \emph {et~al.}(2025)\citenamefont {Valahu}, \citenamefont {Stafford}, \citenamefont {Huang}, \citenamefont {Matsos}, \citenamefont {Millican}, \citenamefont {Chalermpusitarak}, \citenamefont {Menicucci}, \citenamefont {Combes}, \citenamefont {Baragiola},\ and\ \citenamefont {Tan}}]{valahu_sensing_2025}%
  \BibitemOpen
  \bibfield  {author} {\bibinfo {author} {\bibfnamefont {C.~H.}\ \bibnamefont {Valahu}}, \bibinfo {author} {\bibfnamefont {M.~P.}\ \bibnamefont {Stafford}}, \bibinfo {author} {\bibfnamefont {Z.}~\bibnamefont {Huang}}, \bibinfo {author} {\bibfnamefont {V.~G.}\ \bibnamefont {Matsos}}, \bibinfo {author} {\bibfnamefont {M.~J.}\ \bibnamefont {Millican}}, \bibinfo {author} {\bibfnamefont {T.}~\bibnamefont {Chalermpusitarak}}, \bibinfo {author} {\bibfnamefont {N.~C.}\ \bibnamefont {Menicucci}}, \bibinfo {author} {\bibfnamefont {J.}~\bibnamefont {Combes}}, \bibinfo {author} {\bibfnamefont {B.~Q.}\ \bibnamefont {Baragiola}},\ and\ \bibinfo {author} {\bibfnamefont {T.~R.}\ \bibnamefont {Tan}},\ }\bibfield  {title} {\bibinfo {title} {Quantum-enhanced multiparameter sensing in a single mode},\ }\href {https://doi.org/10.1126/sciadv.adw9757} {\bibfield  {journal} {\bibinfo  {journal} {Sci. Adv.}\ }\textbf {\bibinfo {volume} {11}},\ \bibinfo {pages} {eadw9757} (\bibinfo {year} {2025})}\BibitemShut {NoStop}%
\bibitem [{\citenamefont {Chalermpusitarak}\ \emph {et~al.}(2025)\citenamefont {Chalermpusitarak}, \citenamefont {Schwennicke}, \citenamefont {Kassal},\ and\ \citenamefont {Tan}}]{chalermpusitarak2025programmable}%
  \BibitemOpen
  \bibfield  {author} {\bibinfo {author} {\bibfnamefont {T.}~\bibnamefont {Chalermpusitarak}}, \bibinfo {author} {\bibfnamefont {K.}~\bibnamefont {Schwennicke}}, \bibinfo {author} {\bibfnamefont {I.}~\bibnamefont {Kassal}},\ and\ \bibinfo {author} {\bibfnamefont {T.~R.}\ \bibnamefont {Tan}},\ }\href@noop {} {\bibinfo {title} {Programmable generation of arbitrary continuous-variable anharmonicities and nonlinear couplings}} (\bibinfo {year} {2025}),\ \Eprint {https://arxiv.org/abs/2511.22286} {arXiv:2511.22286} \BibitemShut {NoStop}%
\bibitem [{\citenamefont {Hastrup}\ \emph {et~al.}(2021)\citenamefont {Hastrup}, \citenamefont {Larsen}, \citenamefont {Neergaard-Nielsen}, \citenamefont {Menicucci},\ and\ \citenamefont {Andersen}}]{Cubic2021Phase}%
  \BibitemOpen
  \bibfield  {author} {\bibinfo {author} {\bibfnamefont {J.}~\bibnamefont {Hastrup}}, \bibinfo {author} {\bibfnamefont {M.~V.}\ \bibnamefont {Larsen}}, \bibinfo {author} {\bibfnamefont {J.~S.}\ \bibnamefont {Neergaard-Nielsen}}, \bibinfo {author} {\bibfnamefont {N.~C.}\ \bibnamefont {Menicucci}},\ and\ \bibinfo {author} {\bibfnamefont {U.~L.}\ \bibnamefont {Andersen}},\ }\bibfield  {title} {\bibinfo {title} {Unsuitability of cubic phase gates for non-{Clifford} operations on {Gottesman-Kitaev-Preskill} states},\ }\href {https://doi.org/10.1103/PhysRevA.103.032409} {\bibfield  {journal} {\bibinfo  {journal} {Phys. Rev. A}\ }\textbf {\bibinfo {volume} {103}},\ \bibinfo {pages} {032409} (\bibinfo {year} {2021})}\BibitemShut {NoStop}%
\bibitem [{\citenamefont {Krastanov}\ \emph {et~al.}(2015)\citenamefont {Krastanov}, \citenamefont {Albert}, \citenamefont {Shen}, \citenamefont {Zou}, \citenamefont {Heeres}, \citenamefont {Vlastakis}, \citenamefont {Schoelkopf},\ and\ \citenamefont {Jiang}}]{Krastanov2015Universal}%
  \BibitemOpen
  \bibfield  {author} {\bibinfo {author} {\bibfnamefont {S.}~\bibnamefont {Krastanov}}, \bibinfo {author} {\bibfnamefont {V.~V.}\ \bibnamefont {Albert}}, \bibinfo {author} {\bibfnamefont {C.}~\bibnamefont {Shen}}, \bibinfo {author} {\bibfnamefont {C.-L.}\ \bibnamefont {Zou}}, \bibinfo {author} {\bibfnamefont {R.~W.}\ \bibnamefont {Heeres}}, \bibinfo {author} {\bibfnamefont {B.}~\bibnamefont {Vlastakis}}, \bibinfo {author} {\bibfnamefont {R.~J.}\ \bibnamefont {Schoelkopf}},\ and\ \bibinfo {author} {\bibfnamefont {L.}~\bibnamefont {Jiang}},\ }\bibfield  {title} {\bibinfo {title} {Universal control of an oscillator with dispersive coupling to a qubit},\ }\href {https://doi.org/10.1103/PhysRevA.92.040303} {\bibfield  {journal} {\bibinfo  {journal} {Phys. Rev. A}\ }\textbf {\bibinfo {volume} {92}},\ \bibinfo {pages} {040303} (\bibinfo {year} {2015})}\BibitemShut {NoStop}%
\bibitem [{\citenamefont {Diringer}\ \emph {et~al.}(2024)\citenamefont {Diringer}, \citenamefont {Blumenthal}, \citenamefont {Grinberg}, \citenamefont {Jiang},\ and\ \citenamefont {Hacohen-Gourgy}}]{Conditional2024Not}%
  \BibitemOpen
  \bibfield  {author} {\bibinfo {author} {\bibfnamefont {A.~A.}\ \bibnamefont {Diringer}}, \bibinfo {author} {\bibfnamefont {E.}~\bibnamefont {Blumenthal}}, \bibinfo {author} {\bibfnamefont {A.}~\bibnamefont {Grinberg}}, \bibinfo {author} {\bibfnamefont {L.}~\bibnamefont {Jiang}},\ and\ \bibinfo {author} {\bibfnamefont {S.}~\bibnamefont {Hacohen-Gourgy}},\ }\bibfield  {title} {\bibinfo {title} {Conditional-not displacement: Fast multioscillator control with a single qubit},\ }\href {https://doi.org/10.1103/PhysRevX.14.011055} {\bibfield  {journal} {\bibinfo  {journal} {Phys. Rev. X}\ }\textbf {\bibinfo {volume} {14}},\ \bibinfo {pages} {011055} (\bibinfo {year} {2024})}\BibitemShut {NoStop}%
\bibitem [{\citenamefont {Shor}(1995)}]{Shor}%
  \BibitemOpen
  \bibfield  {author} {\bibinfo {author} {\bibfnamefont {P.~W.}\ \bibnamefont {Shor}},\ }\bibfield  {title} {\bibinfo {title} {Scheme for reducing decoherence in quantum computer memory},\ }\href {https://doi.org/10.1103/PhysRevA.52.R2493} {\bibfield  {journal} {\bibinfo  {journal} {Phys. Rev. A}\ }\textbf {\bibinfo {volume} {52}},\ \bibinfo {pages} {R2493} (\bibinfo {year} {1995})}\BibitemShut {NoStop}%
\bibitem [{\citenamefont {Gottesman}(1997)}]{gottesman1997stabilizer}%
  \BibitemOpen
  \bibfield  {author} {\bibinfo {author} {\bibfnamefont {D.}~\bibnamefont {Gottesman}},\ }\emph {\bibinfo {title} {Stabilizer Codes and Quantum Error Correction}},\ \href@noop {} {Ph.D. thesis},\ \bibinfo  {school} {California Institute of Technology} (\bibinfo {year} {1997}),\ \bibinfo {note} {arXiv:quant-ph/9705052}\BibitemShut {NoStop}%
\bibitem [{\citenamefont {Dennis}\ \emph {et~al.}(2002)\citenamefont {Dennis}, \citenamefont {Kitaev}, \citenamefont {Landahl},\ and\ \citenamefont {Preskill}}]{dennis2002topological}%
  \BibitemOpen
  \bibfield  {author} {\bibinfo {author} {\bibfnamefont {E.}~\bibnamefont {Dennis}}, \bibinfo {author} {\bibfnamefont {A.}~\bibnamefont {Kitaev}}, \bibinfo {author} {\bibfnamefont {A.}~\bibnamefont {Landahl}},\ and\ \bibinfo {author} {\bibfnamefont {J.}~\bibnamefont {Preskill}},\ }\bibfield  {title} {\bibinfo {title} {Topological quantum memory},\ }\href {https://doi.org/10.1063/1.1499754} {\bibfield  {journal} {\bibinfo  {journal} {J. Math. Phys.}\ }\textbf {\bibinfo {volume} {43}},\ \bibinfo {pages} {4452} (\bibinfo {year} {2002})}\BibitemShut {NoStop}%
\bibitem [{\citenamefont {Lidar}\ and\ \citenamefont {Brun}(2013)}]{lidar2013quantum}%
  \BibitemOpen
  \bibfield  {author} {\bibinfo {author} {\bibfnamefont {D.~A.}\ \bibnamefont {Lidar}}\ and\ \bibinfo {author} {\bibfnamefont {T.~A.}\ \bibnamefont {Brun}},\ }\href@noop {} {\emph {\bibinfo {title} {Quantum Error Correction}}}\ (\bibinfo  {publisher} {Cambridge University Press},\ \bibinfo {year} {2013})\BibitemShut {NoStop}%
\bibitem [{\citenamefont {Terhal}(2015)}]{Terhal}%
  \BibitemOpen
  \bibfield  {author} {\bibinfo {author} {\bibfnamefont {B.~M.}\ \bibnamefont {Terhal}},\ }\bibfield  {title} {\bibinfo {title} {Quantum error correction for quantum memories},\ }\href {https://doi.org/10.1103/RevModPhys.87.307} {\bibfield  {journal} {\bibinfo  {journal} {Rev. Mod. Phys.}\ }\textbf {\bibinfo {volume} {87}},\ \bibinfo {pages} {307} (\bibinfo {year} {2015})}\BibitemShut {NoStop}%
\bibitem [{\citenamefont {Girvin}(2023)}]{Girvin}%
  \BibitemOpen
  \bibfield  {author} {\bibinfo {author} {\bibfnamefont {S.~M.}\ \bibnamefont {Girvin}},\ }\bibfield  {title} {\bibinfo {title} {Introduction to quantum error correction and fault tolerance},\ }\href {https://doi.org/10.21468/SciPostPhysLectNotes.70} {\bibfield  {journal} {\bibinfo  {journal} {SciPost Phys. Lect. Notes}\ ,\ \bibinfo {pages} {70}} (\bibinfo {year} {2023})}\BibitemShut {NoStop}%
\bibitem [{\citenamefont {Albert}\ and\ \citenamefont {Faist}(2026)}]{ErrorCorrectionZoo}%
  \BibitemOpen
  \bibinfo {editor} {\bibfnamefont {V.~V.}\ \bibnamefont {Albert}}\ and\ \bibinfo {editor} {\bibfnamefont {P.}~\bibnamefont {Faist}},\ eds.,\ \href@noop {} {\emph {\bibinfo {title} {The Error Correction Zoo}}}\ (\bibinfo {year} {2026})\ \bibinfo {note} {\url{https://errorcorrectionzoo.org/}}\BibitemShut {NoStop}%
\bibitem [{\citenamefont {Albert}(2025)}]{Albert2025BosonicCoding}%
  \BibitemOpen
  \bibfield  {author} {\bibinfo {author} {\bibfnamefont {V.~V.}\ \bibnamefont {Albert}},\ }\bibfield  {title} {\bibinfo {title} {Bosonic coding: Introduction and use cases},\ }\href {https://doi.org/10.3254/ENFI250007} {\bibfield  {journal} {\bibinfo  {journal} {Proc. Int. Sch. Phys. ``Enrico Fermi''}\ }\textbf {\bibinfo {volume} {209}},\ \bibinfo {pages} {79} (\bibinfo {year} {2025})}\BibitemShut {NoStop}%
\bibitem [{\citenamefont {Gottesman}\ \emph {et~al.}(2001)\citenamefont {Gottesman}, \citenamefont {Kitaev},\ and\ \citenamefont {Preskill}}]{gottesman2001encoding}%
  \BibitemOpen
  \bibfield  {author} {\bibinfo {author} {\bibfnamefont {D.}~\bibnamefont {Gottesman}}, \bibinfo {author} {\bibfnamefont {A.}~\bibnamefont {Kitaev}},\ and\ \bibinfo {author} {\bibfnamefont {J.}~\bibnamefont {Preskill}},\ }\bibfield  {title} {\bibinfo {title} {Encoding a qubit in an oscillator},\ }\href {https://doi.org/10.1103/PhysRevA.64.012310} {\bibfield  {journal} {\bibinfo  {journal} {Phys. Rev. A}\ }\textbf {\bibinfo {volume} {64}},\ \bibinfo {pages} {012310} (\bibinfo {year} {2001})}\BibitemShut {NoStop}%
\bibitem [{\citenamefont {Leghtas}\ \emph {et~al.}(2013)\citenamefont {Leghtas}, \citenamefont {Kirchmair}, \citenamefont {Vlastakis}, \citenamefont {Schoelkopf}, \citenamefont {Devoret},\ and\ \citenamefont {Mirrahimi}}]{leghtas2013hardware}%
  \BibitemOpen
  \bibfield  {author} {\bibinfo {author} {\bibfnamefont {Z.}~\bibnamefont {Leghtas}}, \bibinfo {author} {\bibfnamefont {G.}~\bibnamefont {Kirchmair}}, \bibinfo {author} {\bibfnamefont {B.}~\bibnamefont {Vlastakis}}, \bibinfo {author} {\bibfnamefont {R.~J.}\ \bibnamefont {Schoelkopf}}, \bibinfo {author} {\bibfnamefont {M.~H.}\ \bibnamefont {Devoret}},\ and\ \bibinfo {author} {\bibfnamefont {M.}~\bibnamefont {Mirrahimi}},\ }\bibfield  {title} {\bibinfo {title} {Hardware-efficient autonomous quantum memory protection},\ }\href {https://doi.org/10.1103/PhysRevLett.111.120501} {\bibfield  {journal} {\bibinfo  {journal} {Phys. Rev. Lett.}\ }\textbf {\bibinfo {volume} {111}},\ \bibinfo {pages} {120501} (\bibinfo {year} {2013})}\BibitemShut {NoStop}%
\bibitem [{\citenamefont {Michael}\ \emph {et~al.}(2016)\citenamefont {Michael}, \citenamefont {Silveri}, \citenamefont {Brierley}, \citenamefont {Albert}, \citenamefont {Salmilehto}, \citenamefont {Jiang},\ and\ \citenamefont {Girvin}}]{michael2016new}%
  \BibitemOpen
  \bibfield  {author} {\bibinfo {author} {\bibfnamefont {M.~H.}\ \bibnamefont {Michael}}, \bibinfo {author} {\bibfnamefont {M.}~\bibnamefont {Silveri}}, \bibinfo {author} {\bibfnamefont {R.~T.}\ \bibnamefont {Brierley}}, \bibinfo {author} {\bibfnamefont {V.~V.}\ \bibnamefont {Albert}}, \bibinfo {author} {\bibfnamefont {J.}~\bibnamefont {Salmilehto}}, \bibinfo {author} {\bibfnamefont {L.}~\bibnamefont {Jiang}},\ and\ \bibinfo {author} {\bibfnamefont {S.~M.}\ \bibnamefont {Girvin}},\ }\bibfield  {title} {\bibinfo {title} {New class of quantum error-correcting codes for a bosonic mode},\ }\href {https://doi.org/10.1103/PhysRevX.6.031006} {\bibfield  {journal} {\bibinfo  {journal} {Phys. Rev. X}\ }\textbf {\bibinfo {volume} {6}},\ \bibinfo {pages} {031006} (\bibinfo {year} {2016})}\BibitemShut {NoStop}%
\bibitem [{\citenamefont {Niu}\ \emph {et~al.}(2018)\citenamefont {Niu}, \citenamefont {Chuang},\ and\ \citenamefont {Shapiro}}]{PhysRevA.97.032323}%
  \BibitemOpen
  \bibfield  {author} {\bibinfo {author} {\bibfnamefont {M.~Y.}\ \bibnamefont {Niu}}, \bibinfo {author} {\bibfnamefont {I.~L.}\ \bibnamefont {Chuang}},\ and\ \bibinfo {author} {\bibfnamefont {J.~H.}\ \bibnamefont {Shapiro}},\ }\bibfield  {title} {\bibinfo {title} {Hardware-efficient bosonic quantum error-correcting codes based on symmetry operators},\ }\href {https://doi.org/10.1103/PhysRevA.97.032323} {\bibfield  {journal} {\bibinfo  {journal} {Phys. Rev. A}\ }\textbf {\bibinfo {volume} {97}},\ \bibinfo {pages} {032323} (\bibinfo {year} {2018})}\BibitemShut {NoStop}%
\bibitem [{\citenamefont {Grimsmo}\ \emph {et~al.}(2020)\citenamefont {Grimsmo}, \citenamefont {Combes},\ and\ \citenamefont {Baragiola}}]{grimsmo2020quantum}%
  \BibitemOpen
  \bibfield  {author} {\bibinfo {author} {\bibfnamefont {A.~L.}\ \bibnamefont {Grimsmo}}, \bibinfo {author} {\bibfnamefont {J.}~\bibnamefont {Combes}},\ and\ \bibinfo {author} {\bibfnamefont {B.~Q.}\ \bibnamefont {Baragiola}},\ }\bibfield  {title} {\bibinfo {title} {Quantum computing with rotation-symmetric bosonic codes},\ }\href {https://doi.org/10.1103/PhysRevX.10.011058} {\bibfield  {journal} {\bibinfo  {journal} {Phys. Rev. X}\ }\textbf {\bibinfo {volume} {10}},\ \bibinfo {pages} {011058} (\bibinfo {year} {2020})}\BibitemShut {NoStop}%
\bibitem [{\citenamefont {Bartlett}\ \emph {et~al.}(2002)\citenamefont {Bartlett}, \citenamefont {de~Guise},\ and\ \citenamefont {Sanders}}]{PhysRevA.65.052316}%
  \BibitemOpen
  \bibfield  {author} {\bibinfo {author} {\bibfnamefont {S.~D.}\ \bibnamefont {Bartlett}}, \bibinfo {author} {\bibfnamefont {H.}~\bibnamefont {de~Guise}},\ and\ \bibinfo {author} {\bibfnamefont {B.~C.}\ \bibnamefont {Sanders}},\ }\bibfield  {title} {\bibinfo {title} {Quantum encodings in spin systems and harmonic oscillators},\ }\href {https://doi.org/10.1103/PhysRevA.65.052316} {\bibfield  {journal} {\bibinfo  {journal} {Phys. Rev. A}\ }\textbf {\bibinfo {volume} {65}},\ \bibinfo {pages} {052316} (\bibinfo {year} {2002})}\BibitemShut {NoStop}%
\bibitem [{\citenamefont {Wu}\ \emph {et~al.}(2023)\citenamefont {Wu}, \citenamefont {Brady},\ and\ \citenamefont {Zhuang}}]{wu2023optimal}%
  \BibitemOpen
  \bibfield  {author} {\bibinfo {author} {\bibfnamefont {J.}~\bibnamefont {Wu}}, \bibinfo {author} {\bibfnamefont {A.~J.}\ \bibnamefont {Brady}},\ and\ \bibinfo {author} {\bibfnamefont {Q.}~\bibnamefont {Zhuang}},\ }\bibfield  {title} {\bibinfo {title} {Optimal encoding of oscillators into more oscillators},\ }\href {https://doi.org/10.22331/q-2023-08-16-1082} {\bibfield  {journal} {\bibinfo  {journal} {Quantum}\ }\textbf {\bibinfo {volume} {7}},\ \bibinfo {pages} {1082} (\bibinfo {year} {2023})}\BibitemShut {NoStop}%
\bibitem [{\citenamefont {Xu}\ \emph {et~al.}(2024)\citenamefont {Xu}, \citenamefont {Wang},\ and\ \citenamefont {Albert}}]{PhysRevA.110.022402}%
  \BibitemOpen
  \bibfield  {author} {\bibinfo {author} {\bibfnamefont {Y.}~\bibnamefont {Xu}}, \bibinfo {author} {\bibfnamefont {Y.}~\bibnamefont {Wang}},\ and\ \bibinfo {author} {\bibfnamefont {V.~V.}\ \bibnamefont {Albert}},\ }\bibfield  {title} {\bibinfo {title} {Multimode rotation-symmetric bosonic codes from homological rotor codes},\ }\href {https://doi.org/10.1103/PhysRevA.110.022402} {\bibfield  {journal} {\bibinfo  {journal} {Phys. Rev. A}\ }\textbf {\bibinfo {volume} {110}},\ \bibinfo {pages} {022402} (\bibinfo {year} {2024})}\BibitemShut {NoStop}%
\bibitem [{\citenamefont {Terhal}\ \emph {et~al.}(2020)\citenamefont {Terhal}, \citenamefont {Conrad},\ and\ \citenamefont {Vuillot}}]{terhal2020towards}%
  \BibitemOpen
  \bibfield  {author} {\bibinfo {author} {\bibfnamefont {B.~M.}\ \bibnamefont {Terhal}}, \bibinfo {author} {\bibfnamefont {J.}~\bibnamefont {Conrad}},\ and\ \bibinfo {author} {\bibfnamefont {C.}~\bibnamefont {Vuillot}},\ }\bibfield  {title} {\bibinfo {title} {Towards scalable bosonic quantum error correction},\ }\href {https://doi.org/10.1088/2058-9565/ab98a5} {\bibfield  {journal} {\bibinfo  {journal} {Quantum Sci. Technol.}\ }\textbf {\bibinfo {volume} {5}},\ \bibinfo {pages} {043001} (\bibinfo {year} {2020})}\BibitemShut {NoStop}%
\bibitem [{\citenamefont {Xu}\ \emph {et~al.}(2023)\citenamefont {Xu}, \citenamefont {Wang}, \citenamefont {Kuo},\ and\ \citenamefont {Albert}}]{Qubit_Oscillator_Concatenated}%
  \BibitemOpen
  \bibfield  {author} {\bibinfo {author} {\bibfnamefont {Y.}~\bibnamefont {Xu}}, \bibinfo {author} {\bibfnamefont {Y.}~\bibnamefont {Wang}}, \bibinfo {author} {\bibfnamefont {E.-J.}\ \bibnamefont {Kuo}},\ and\ \bibinfo {author} {\bibfnamefont {V.~V.}\ \bibnamefont {Albert}},\ }\bibfield  {title} {\bibinfo {title} {Qubit-oscillator concatenated codes: Decoding formalism and code comparison},\ }\href {https://doi.org/10.1103/PRXQuantum.4.020342} {\bibfield  {journal} {\bibinfo  {journal} {PRX Quantum}\ }\textbf {\bibinfo {volume} {4}},\ \bibinfo {pages} {020342} (\bibinfo {year} {2023})}\BibitemShut {NoStop}%
\bibitem [{\citenamefont {Brock}\ \emph {et~al.}(2025)\citenamefont {Brock}, \citenamefont {Singh}, \citenamefont {Eickbusch}, \citenamefont {Sivak}, \citenamefont {Ding}, \citenamefont {Frunzio}, \citenamefont {Girvin},\ and\ \citenamefont {Devoret}}]{brock_quantum_2025}%
  \BibitemOpen
  \bibfield  {author} {\bibinfo {author} {\bibfnamefont {B.~L.}\ \bibnamefont {Brock}}, \bibinfo {author} {\bibfnamefont {S.}~\bibnamefont {Singh}}, \bibinfo {author} {\bibfnamefont {A.}~\bibnamefont {Eickbusch}}, \bibinfo {author} {\bibfnamefont {V.~V.}\ \bibnamefont {Sivak}}, \bibinfo {author} {\bibfnamefont {A.~Z.}\ \bibnamefont {Ding}}, \bibinfo {author} {\bibfnamefont {L.}~\bibnamefont {Frunzio}}, \bibinfo {author} {\bibfnamefont {S.~M.}\ \bibnamefont {Girvin}},\ and\ \bibinfo {author} {\bibfnamefont {M.~H.}\ \bibnamefont {Devoret}},\ }\bibfield  {title} {\bibinfo {title} {Quantum error correction of qudits beyond break-even},\ }\href {https://doi.org/10.1038/s41586-025-08899-y} {\bibfield  {journal} {\bibinfo  {journal} {Nature}\ }\textbf {\bibinfo {volume} {641}},\ \bibinfo {pages} {612} (\bibinfo {year} {2025})}\BibitemShut {NoStop}%
\bibitem [{\citenamefont {Joshi}\ \emph {et~al.}(2021)\citenamefont {Joshi}, \citenamefont {Noh},\ and\ \citenamefont {Gao}}]{joshi2021quantum}%
  \BibitemOpen
  \bibfield  {author} {\bibinfo {author} {\bibfnamefont {A.}~\bibnamefont {Joshi}}, \bibinfo {author} {\bibfnamefont {K.}~\bibnamefont {Noh}},\ and\ \bibinfo {author} {\bibfnamefont {Y.~Y.}\ \bibnamefont {Gao}},\ }\bibfield  {title} {\bibinfo {title} {Quantum information processing with bosonic qubits in circuit {QED}},\ }\href {https://doi.org/10.1088/2058-9565/abe989} {\bibfield  {journal} {\bibinfo  {journal} {Quantum Sci. Technol.}\ }\textbf {\bibinfo {volume} {6}},\ \bibinfo {pages} {033001} (\bibinfo {year} {2021})}\BibitemShut {NoStop}%
\bibitem [{\citenamefont {Cai}\ \emph {et~al.}(2021)\citenamefont {Cai}, \citenamefont {Ma}, \citenamefont {Wang}, \citenamefont {Zou},\ and\ \citenamefont {Sun}}]{cai_bosonic_2021}%
  \BibitemOpen
  \bibfield  {author} {\bibinfo {author} {\bibfnamefont {W.}~\bibnamefont {Cai}}, \bibinfo {author} {\bibfnamefont {Y.}~\bibnamefont {Ma}}, \bibinfo {author} {\bibfnamefont {W.}~\bibnamefont {Wang}}, \bibinfo {author} {\bibfnamefont {C.-L.}\ \bibnamefont {Zou}},\ and\ \bibinfo {author} {\bibfnamefont {L.}~\bibnamefont {Sun}},\ }\bibfield  {title} {\bibinfo {title} {Bosonic quantum error correction codes in superconducting quantum circuits},\ }\href {https://doi.org/10.1016/j.fmre.2020.12.006} {\bibfield  {journal} {\bibinfo  {journal} {Fundam. Res.}\ }\textbf {\bibinfo {volume} {1}},\ \bibinfo {pages} {50} (\bibinfo {year} {2021})}\BibitemShut {NoStop}%
\bibitem [{\citenamefont {H{\"a}nggli}\ and\ \citenamefont {K{\"o}nig}(2022)}]{hanggli2022oscillator}%
  \BibitemOpen
  \bibfield  {author} {\bibinfo {author} {\bibfnamefont {L.}~\bibnamefont {H{\"a}nggli}}\ and\ \bibinfo {author} {\bibfnamefont {R.}~\bibnamefont {K{\"o}nig}},\ }\bibfield  {title} {\bibinfo {title} {Oscillator-to-oscillator codes do not have a threshold},\ }\href {https://doi.org/10.1109/TIT.2021.3126881} {\bibfield  {journal} {\bibinfo  {journal} {IEEE Trans. Inf. Theory}\ }\textbf {\bibinfo {volume} {68}},\ \bibinfo {pages} {1068} (\bibinfo {year} {2022})}\BibitemShut {NoStop}%
\bibitem [{\citenamefont {Noh}\ \emph {et~al.}(2020)\citenamefont {Noh}, \citenamefont {Girvin},\ and\ \citenamefont {Jiang}}]{noh2020encodingar}%
  \BibitemOpen
  \bibfield  {author} {\bibinfo {author} {\bibfnamefont {K.}~\bibnamefont {Noh}}, \bibinfo {author} {\bibfnamefont {S.~M.}\ \bibnamefont {Girvin}},\ and\ \bibinfo {author} {\bibfnamefont {L.}~\bibnamefont {Jiang}},\ }\bibfield  {title} {\bibinfo {title} {Encoding an oscillator into many oscillators},\ }\href {https://doi.org/10.1103/PhysRevLett.125.080503} {\bibfield  {journal} {\bibinfo  {journal} {Phys. Rev. Lett.}\ }\textbf {\bibinfo {volume} {125}},\ \bibinfo {pages} {080503} (\bibinfo {year} {2020})},\ \bibinfo {note} {the preprint arXiv:1903.12615 contains additional material not included in the published version, including an implementation of the logical Gaussian operations.}\BibitemShut {Stop}%
\bibitem [{\citenamefont {Brady}\ \emph {et~al.}(2024)\citenamefont {Brady}, \citenamefont {Wu},\ and\ \citenamefont {Zhuang}}]{brady2024safeguarding}%
  \BibitemOpen
  \bibfield  {author} {\bibinfo {author} {\bibfnamefont {A.~J.}\ \bibnamefont {Brady}}, \bibinfo {author} {\bibfnamefont {J.}~\bibnamefont {Wu}},\ and\ \bibinfo {author} {\bibfnamefont {Q.}~\bibnamefont {Zhuang}},\ }\bibfield  {title} {\bibinfo {title} {Safeguarding oscillators and qudits with distributed two-mode squeezing},\ }\href {https://doi.org/10.22331/q-2024-09-19-1478} {\bibfield  {journal} {\bibinfo  {journal} {Quantum}\ }\textbf {\bibinfo {volume} {8}},\ \bibinfo {pages} {1478} (\bibinfo {year} {2024})}\BibitemShut {NoStop}%
\bibitem [{\citenamefont {Niset}\ \emph {et~al.}(2009)\citenamefont {Niset}, \citenamefont {Fiur\'a\ifmmode~\check{s}\else \v{s}\fi{}ek},\ and\ \citenamefont {Cerf}}]{PhysRevLett.102.120501}%
  \BibitemOpen
  \bibfield  {author} {\bibinfo {author} {\bibfnamefont {J.}~\bibnamefont {Niset}}, \bibinfo {author} {\bibfnamefont {J.}~\bibnamefont {Fiur\'a\ifmmode~\check{s}\else \v{s}\fi{}ek}},\ and\ \bibinfo {author} {\bibfnamefont {N.~J.}\ \bibnamefont {Cerf}},\ }\bibfield  {title} {\bibinfo {title} {No-go theorem for {Gaussian} quantum error correction},\ }\href {https://doi.org/10.1103/PhysRevLett.102.120501} {\bibfield  {journal} {\bibinfo  {journal} {Phys. Rev. Lett.}\ }\textbf {\bibinfo {volume} {102}},\ \bibinfo {pages} {120501} (\bibinfo {year} {2009})}\BibitemShut {NoStop}%
\bibitem [{\citenamefont {Eisert}\ \emph {et~al.}(2002)\citenamefont {Eisert}, \citenamefont {Scheel},\ and\ \citenamefont {Plenio}}]{Distilling_Impossible}%
  \BibitemOpen
  \bibfield  {author} {\bibinfo {author} {\bibfnamefont {J.}~\bibnamefont {Eisert}}, \bibinfo {author} {\bibfnamefont {S.}~\bibnamefont {Scheel}},\ and\ \bibinfo {author} {\bibfnamefont {M.~B.}\ \bibnamefont {Plenio}},\ }\bibfield  {title} {\bibinfo {title} {Distilling {Gaussian} states with {Gaussian} operations is impossible},\ }\href {https://doi.org/10.1103/PhysRevLett.89.137903} {\bibfield  {journal} {\bibinfo  {journal} {Phys. Rev. Lett.}\ }\textbf {\bibinfo {volume} {89}},\ \bibinfo {pages} {137903} (\bibinfo {year} {2002})}\BibitemShut {NoStop}%
\bibitem [{\citenamefont {Sutherland}\ and\ \citenamefont {Srinivas}(2021)}]{sutherland2021universal}%
  \BibitemOpen
  \bibfield  {author} {\bibinfo {author} {\bibfnamefont {R.~T.}\ \bibnamefont {Sutherland}}\ and\ \bibinfo {author} {\bibfnamefont {R.}~\bibnamefont {Srinivas}},\ }\bibfield  {title} {\bibinfo {title} {Universal hybrid quantum computing in trapped ions},\ }\href {https://doi.org/10.1103/PhysRevA.104.032609} {\bibfield  {journal} {\bibinfo  {journal} {Phys. Rev. A}\ }\textbf {\bibinfo {volume} {104}},\ \bibinfo {pages} {032609} (\bibinfo {year} {2021})}\BibitemShut {NoStop}%
\bibitem [{\citenamefont {Eickbusch}\ \emph {et~al.}(2022)\citenamefont {Eickbusch}, \citenamefont {Sivak}, \citenamefont {Ding}, \citenamefont {Elder}, \citenamefont {Jha}, \citenamefont {Venkatraman}, \citenamefont {Royer}, \citenamefont {Girvin}, \citenamefont {Schoelkopf},\ and\ \citenamefont {Devoret}}]{eickbusch2022fast}%
  \BibitemOpen
  \bibfield  {author} {\bibinfo {author} {\bibfnamefont {A.}~\bibnamefont {Eickbusch}}, \bibinfo {author} {\bibfnamefont {V.}~\bibnamefont {Sivak}}, \bibinfo {author} {\bibfnamefont {A.~Z.}\ \bibnamefont {Ding}}, \bibinfo {author} {\bibfnamefont {S.~S.}\ \bibnamefont {Elder}}, \bibinfo {author} {\bibfnamefont {S.~R.}\ \bibnamefont {Jha}}, \bibinfo {author} {\bibfnamefont {J.}~\bibnamefont {Venkatraman}}, \bibinfo {author} {\bibfnamefont {B.}~\bibnamefont {Royer}}, \bibinfo {author} {\bibfnamefont {S.~M.}\ \bibnamefont {Girvin}}, \bibinfo {author} {\bibfnamefont {R.~J.}\ \bibnamefont {Schoelkopf}},\ and\ \bibinfo {author} {\bibfnamefont {M.~H.}\ \bibnamefont {Devoret}},\ }\bibfield  {title} {\bibinfo {title} {Fast universal control of an oscillator with weak dispersive coupling to a qubit},\ }\href {https://doi.org/10.1038/s41567-022-01776-9} {\bibfield  {journal} {\bibinfo  {journal} {Nat. Phys.}\ }\textbf {\bibinfo {volume} {18}},\ \bibinfo {pages} {1464} (\bibinfo {year} {2022})}\BibitemShut {NoStop}%
\bibitem [{\citenamefont {Kang}\ \emph {et~al.}(2025)\citenamefont {Kang}, \citenamefont {Soley}, \citenamefont {Crane}, \citenamefont {Girvin},\ and\ \citenamefont {Wiebe}}]{kang_leveraging_2025}%
  \BibitemOpen
  \bibfield  {author} {\bibinfo {author} {\bibfnamefont {C.}~\bibnamefont {Kang}}, \bibinfo {author} {\bibfnamefont {M.~B.}\ \bibnamefont {Soley}}, \bibinfo {author} {\bibfnamefont {E.}~\bibnamefont {Crane}}, \bibinfo {author} {\bibfnamefont {S.~M.}\ \bibnamefont {Girvin}},\ and\ \bibinfo {author} {\bibfnamefont {N.}~\bibnamefont {Wiebe}},\ }\bibfield  {title} {\bibinfo {title} {Leveraging {Hamiltonian} simulation techniques to compile operations on bosonic devices},\ }\href {https://doi.org/10.1088/1751-8121/adb5df} {\bibfield  {journal} {\bibinfo  {journal} {J. Phys. A: Math. Theor.}\ }\textbf {\bibinfo {volume} {58}},\ \bibinfo {pages} {175301} (\bibinfo {year} {2025})}\BibitemShut {NoStop}%
\bibitem [{\citenamefont {Glancy}\ and\ \citenamefont {Knill}(2006)}]{Knill_GKP}%
  \BibitemOpen
  \bibfield  {author} {\bibinfo {author} {\bibfnamefont {S.}~\bibnamefont {Glancy}}\ and\ \bibinfo {author} {\bibfnamefont {E.}~\bibnamefont {Knill}},\ }\bibfield  {title} {\bibinfo {title} {Error analysis for encoding a qubit in an oscillator},\ }\href {https://doi.org/10.1103/PhysRevA.73.012325} {\bibfield  {journal} {\bibinfo  {journal} {Phys. Rev. A}\ }\textbf {\bibinfo {volume} {73}},\ \bibinfo {pages} {012325} (\bibinfo {year} {2006})}\BibitemShut {NoStop}%
\bibitem [{\citenamefont {Noh}\ \emph {et~al.}(2019)\citenamefont {Noh}, \citenamefont {Albert},\ and\ \citenamefont {Jiang}}]{noh2018quantum}%
  \BibitemOpen
  \bibfield  {author} {\bibinfo {author} {\bibfnamefont {K.}~\bibnamefont {Noh}}, \bibinfo {author} {\bibfnamefont {V.~V.}\ \bibnamefont {Albert}},\ and\ \bibinfo {author} {\bibfnamefont {L.}~\bibnamefont {Jiang}},\ }\bibfield  {title} {\bibinfo {title} {Quantum capacity bounds of {Gaussian} thermal loss channels and achievable rates with {Gottesman-Kitaev-Preskill} codes},\ }\href {https://doi.org/10.1109/TIT.2018.2873764} {\bibfield  {journal} {\bibinfo  {journal} {IEEE Trans. Inf. Theory}\ }\textbf {\bibinfo {volume} {65}},\ \bibinfo {pages} {2563} (\bibinfo {year} {2019})}\BibitemShut {NoStop}%
\bibitem [{\citenamefont {Girvin}(2019)}]{girvin2019schrodinger}%
  \BibitemOpen
  \bibfield  {author} {\bibinfo {author} {\bibfnamefont {S.~M.}\ \bibnamefont {Girvin}},\ }\bibfield  {title} {\bibinfo {title} {Schr{\"o}dinger cat states in circuit {QED}},\ }in\ \href {https://doi.org/10.1093/oso/9780198837190.003.0011} {\emph {\bibinfo {booktitle} {Current Trends in Atomic Physics}}},\ \bibinfo {editor} {edited by\ \bibinfo {editor} {\bibfnamefont {A.}~\bibnamefont {Browaeys}}, \bibinfo {editor} {\bibfnamefont {T.}~\bibnamefont {Lahaye}}, \bibinfo {editor} {\bibfnamefont {T.}~\bibnamefont {Porto}}, \bibinfo {editor} {\bibfnamefont {C.~S.}\ \bibnamefont {Adams}}, \bibinfo {editor} {\bibfnamefont {M.}~\bibnamefont {Weidem{\"u}ller}},\ and\ \bibinfo {editor} {\bibfnamefont {L.~F.}\ \bibnamefont {Cugliandolo}}}\ (\bibinfo  {publisher} {Oxford University Press},\ \bibinfo {year} {2019})\ pp.\ \bibinfo {pages} {402--427}\BibitemShut {NoStop}%
\bibitem [{\citenamefont {Kay}(1993)}]{Kay1993}%
  \BibitemOpen
  \bibfield  {author} {\bibinfo {author} {\bibfnamefont {S.~M.}\ \bibnamefont {Kay}},\ }\href@noop {} {\emph {\bibinfo {title} {Fundamentals of Statistical Signal Processing, Volume I: Estimation Theory}}}\ (\bibinfo  {publisher} {Prentice Hall},\ \bibinfo {year} {1993})\BibitemShut {NoStop}%
\bibitem [{\citenamefont {Chakraborty}\ and\ \citenamefont {Albert}(2026)}]{Sayan}%
  \BibitemOpen
  \bibfield  {author} {\bibinfo {author} {\bibfnamefont {S.}~\bibnamefont {Chakraborty}}\ and\ \bibinfo {author} {\bibfnamefont {V.~V.}\ \bibnamefont {Albert}},\ }\bibfield  {title} {\bibinfo {title} {Hybrid oscillator-qudit quantum processors: Stabilizer states, stabilizer codes, symplectic operations, and noncommutative geometry},\ }\href {https://doi.org/10.1103/yr64-ypxt} {\bibfield  {journal} {\bibinfo  {journal} {PRX Quantum}\ ,\ \bibinfo {pages} {doi:10.1103/yr64}} (\bibinfo {year} {2026})}\BibitemShut {NoStop}%
\end{thebibliography}%

\end{document}